\documentclass[prd, preprint, nofootinbib]{revtex4}
\usepackage[dvipdfmx]{graphicx}
\usepackage{amsfonts}
\usepackage{latexsym}
\usepackage{amssymb}
\usepackage{epsfig}
\usepackage{comment}
\usepackage{here}	
\usepackage{amsmath}

\def\slashb#1{\not\!\!#1}
\newcommand{\im}[1]{\text{Im}\,#1}
\newcommand{\re}[1]{\text{Re}\,#1}

\begin{document}

\title{Zero-modes on orbifolds: \\ 
magnetized orbifold models by modular transformation}

\author{Tatsuo Kobayashi, and Satoshi Nagamoto}
 \affiliation{
Department of Physics, Hokkaido University, Sapporo 060-0810, Japan}



\begin{abstract}
We study $T^2/Z_N$ orbifold models with magnetic fluxes.
We propose a systematic way to analyze the number of zero-modes and their wavefunctions 
by use of modular transformation.
Our results are consistent with the previous results, and 
our approach is more direct and analytical than the previous ones.
The index theorem implies that the zero-mode number of the Dirac operator on $T^2$ is equal to the 
index $M$, which corresponds to the magnetic flux in a certain unit.
Our results show that the zero-mode number of the Dirac operator on $T^2/Z_N$ 
is equal to $\lfloor M/N \rfloor +1$ except one case on the $T^2/Z_3$ orbifold.
\end{abstract}

\preprint{EPHOU-17-013}
\preprint{}

\vspace*{3cm}
\maketitle



\section{Introduction}

Superstring theory is a promising candidate for unified theory including gravity, and 
leads to six dimensional space in addition to our four-dimensional (4D) spacetime.
Thus, extra dimensional models are well-motivated.
Indeed, many studies have been carried out.
It is a key point how to derive 4D chiral theory starting from extra dimensional theories, 
because the standard model is a chiral theory.
For example, the toroidal compactification is one of the simplest compactifications, but 
leads to non-chiral theory.
Then, the simple toroidal compactification is not realistic.
However, the torus compactification with magnetic fluxes can lead to 
4D chiral theory from extra dimensional theories as well as 
superstring theories \cite{Bachas:1995ik,Berkooz:1996km,Blumenhagen:2000wh,Angelantonj:2000hi}.
In addition, magnitude of magnetic flux determines the number of zero-modes, 
which would correspond to the generation number.
Also zero-mode profiles are quasi-localized and can lead to suppressed couplings depending 
on their localized points.
Hence, the torus compactification with magnetic fluxes is quite interesting.
Indeed, several studies have been done, e.g. on computation of Yukawa couplings \cite{Cremades:2004wa}, 
higher order couplings \cite{Abe:2009dr}, non-Abelian flavor symmetries \cite{Abe:2009vi,BerasaluceGonzalez:2012vb}, 
massive modes and their phenomenological effects \cite{BerasaluceGonzalez:2012vb,Hamada:2012wj,Buchmuller:2016gib,Ghilencea:2017jmh}, 
etc.\footnote{See also \cite{Abe:2015mua,Abe:2015xua}.}

The orbifold models with magnetic fluxes are also interesting.
Orbifolding can project out the adjoint matter fields corresponding to open string moduli, which 
remain massless in the toroidal compactification with magnetic fluxes.
The number of zero-modes and their profiles in orbifold models are different from 
those in toroidal models \cite{Abe:2008fi}.
Thus, orbifold models with magnetic fluxes have rich structures in model building.
Indeed, $Z_2$ orbifold models have been studied on several aspects, e.g. 
model building \cite{Abe:2008sx,Abe:2012ya,Abe:2016jsb,Abe:2016zgq,Abe:2017gye}, realization of quark and lepton masses 
and their mixing angles and CP phase \cite{Abe:2012fj,Abe:2014vza,Kobayashi:2015siy,Kobayashi:2016qag}.
In addition, it is possible to introduce some degree of freedom on orbifold fixed points, e.g. 
localized modes and localized operators.
That makes phenomenological aspects richer \cite{Buchmuller:2015eya,Buchmuller:2015jna,Buchmuller:2017vho,Ishida:2017avx}.

Other $Z_N$ orbifold models with $N=3,4,6$ have been also studied.
Zero-mode wavefunctions were studied by numerical studies \cite{Abe:2013bca} and 
the corresponding states were studied by operator analysis in quantum mechanism \cite{Abe:2014noa}.
By use of those results, model building and fermion mass matrices were also studied \cite{Abe:2015yva,Matsumoto:2016okl,Fujimoto:2016zjs}.
However, the numerical study is not analytical and results from both approaches were rather complicated.
Simpler approach would be useful for further applications.

Here, we study $Z_N$ orbifold models with magnetic fluxes.
In particular we study the number of zero-modes and their wavefunctions directly 
by using modular transformation.
The modular transformation is a geometrical transformation of the lattice which is used to 
construct $T^2$.
Zero-mode wavefunctions can be written in terms of theta functions, 
which have characteristic behavior under modular transformation.
When we fix a value of complex structure properly, certain modular 
transformation behaves as $Z_N$ twists with $N=3,4,6$.
Using such behavior, we can obtain zero-mode wave functions 
on $Z_N$ orbifolds.
For generic values of magnetic flux, we compute the number of zero-modes 
with each $Z_N$ eigenvalue on $T^2/Z_N$.
We show that the number of $Z_N$ invariant zero-modes is almost universal 
on different $T^2/Z_N$ orbifolds, 
and it is equal to $\lfloor M/N \rfloor +1$ for magnetic flux $M$ in a certain unit 
except one case in the $T^2/Z_3$ orbifold, 
where $\lfloor r \rfloor$ denotes the maximum integer $n$ satisfying $n \leq r$.
Alternatively, the number of $Z_3$ invariant zero-modes is written by 
    $2\lfloor M/(2N) \rfloor +1$ .

This paper is organized as follows.
In section \ref{sec:2}, we review wavefunctions on the two-dimension torus $T^2$ 
with magnetic fluxes as well as the $T^2/Z_2$ orbifold.
In section \ref{sec:3}, we study the $T^2/Z_4$ orbifold.
In section \ref{sec:4} we study the $T^2/Z_3$ orbifold as well as $T^2/Z_6$ orbifold.
In section \ref{sec:5}, we give a comment on our universal result on the number of $Z_N$ invariant zero-modes. 
Section \ref{sec:conclusion} is conclusion.
In Appendix \ref{sec:app-1}, we show computations on the normalization factor 
of zero-mode wave function and inner product of two types of wavefunctions.
Such computations are useful for Sections \ref{sec:3} and \ref{sec:4}.
In Appendix \ref{sec:app-2}, we show the computation on products of the $Z_3$ matrix.
In Appendix \ref{sec:app-3}, we show explicitly zero-mode wavefunctions on the $T^2/Z_4$ orbifold.

\section{Torus model with magnetic flux}
\label{sec:2}

Our starting point is the gauge theory with $2n$ extra dimensions, which are chosen as 
$(T^2)^n$.
Our theory includes the spinor field $\lambda$  and its 
Lagrangian is written by 
\begin{equation}
{\cal L} = -\frac{1}{4g^2}{\rm Tr}F^{MN}F_{FM} -\frac{i}{2g^2}\bar \lambda^MD_M \lambda,
\end{equation}
where $F_{MN} = \partial_M A_N - \partial_N A_M$.
Here, we set the kinetic term of $\lambda$ as one in super Yang-Mills theory, because 
we are motivated from such a theory.
For simplicity, we concentrate on U(1) gauge theory with $n=1$ and spinor field with charge $q$.
Similarly, we can extend our analysis to non-Abelian gauge theory with $n\geq 1$.

We decompose 
\begin{equation}
\lambda(x^\mu,y^m) = \sum_n \eta_n(x^\mu) \otimes \psi_n(y^m),
\end{equation}
where $x^\mu$ denotes coordinates of four dimensional spacetime, while 
$y^m$ with $m=1,2$ denotes coordinates on $T^2$.
$\psi_n(y^m)$ are eigenfunctions of Dirac operator on $T^2$.
In what follows, we concentrate on the zero-modes,  $\psi_0(y)$, which correspond to 
massless modes in 4D effective field theory, and we denote them by 
$\psi(y)$.

\subsection{Magnetized torus models}

Here, we give a review on the $T^2$ model with magnetic flux, in particular zero-mode 
wavefunctions \cite{Cremades:2004wa}.
We use the complex coordinate $z=y^1 + \tau y^2$ instead of the real coordinates, $y^1$ and $y^2$, 
where $\tau$ is a complex, and the metric is given as 
$ds^2 = g_{\alpha \beta}dz^{\alpha} d\bar{z}^{\beta} $,
\begin{equation}
g_{\alpha \beta} = \left(
\begin{array}{cc}
g_{zz} & g_{z \bar{z}} \\
g_{\bar{z} z} & g_{\bar{z} \bar{z}}
\end{array}
\right) = (2\pi R)^2 \left(
\begin{array}{cc}
0 & \frac{1}{2} \\
\frac{1}{2} & 0
\end{array}
\right) .
\end{equation}
To realize the $T^2$, we identify $z \sim z +1$ and $z \sim z + \tau$.

We consider the U(1) magnetic flux $F$ on $T^2$, 
\begin{equation}
F = i\frac{\pi M}{\im{\tau}}  (dz \wedge d\bar{z}).
\end{equation}

Such a magnetic flux can be obtained from the following vector potential,
\begin{equation}
A(z) = \frac{\pi M}{\im{\tau}} \im{(\bar{z}dz)}.
\end{equation}
It satisfies the boundary conditions,
\begin{equation}
\label{eq:A-bc}
A(z+1) = A(z) + d \phi_1, \qquad
A(z+\tau) = A(z) + d \phi_2,
\end{equation}
where
\begin{equation}
\phi_1 = \frac{\pi M}{\im{\tau}} \im{z}, \qquad
\phi_2 =  \frac{\pi M}{\im{\tau}} \im{\bar{\tau}z} .
\end{equation}

Now, let us study the spinor field with U(1) charge $q$ on $T^2$,
\begin{equation}
\psi(z,\bar{z}) = \left(
\begin{array}{c}
\psi_+ \\
\psi_- 
\end{array}
\right).
\end{equation}
We use the gamma matrices, 
\begin{equation}
\Gamma^z = (2\pi R)^{-1} \left(
\begin{array}{cc}
0 & 2 \\
0 & 0 
\end{array}
\right), \ \ \  \Gamma^{\bar{z}} = (2\pi R)^{-1} \left(
\begin{array}{cc}
0 & 0 \\
2 & 0 
\end{array}
\right).
\end{equation}
Then, the Dirac operator on $\psi$ is written by 
\begin{equation}
i \slashb{D} = i \Gamma^z \nabla_z + i \Gamma^{\bar{z}} \nabla_{\bar{z}} 
= \frac{i}{\pi R} \left(
\begin{array}{cc}
0 & D^{\dagger} \\
D  &0
\end{array}
\right),
\end{equation}
where 
\begin{equation}
D^{\dagger} \equiv \partial - q \frac{\pi M}{2 \im{\tau}} \bar{z}, \qquad
D \equiv \bar{\partial} + q \frac{\pi M}{2 \im{\tau}} z .
\end{equation}
Thus, the zero mode equations of spinor are written by 
\begin{equation}
D \psi_+ = 0, \ \ \ D^{\dagger} \psi_- =0.
\end{equation}
Also, they must satisfy the following boundary condition, 
\begin{eqnarray}
\label{eq:z+1}
\psi_\pm(z+1) &=& e^{iq\phi_1(z)}\psi_\pm(z) = \exp \left\{ i \frac{\pi qM}{\im{\tau}} \im{z} \right\} \psi_\pm(z), \\
\label{eq:z+tau}
\psi_\pm(z + \tau) &=& e^{iq\phi_2(z)}\psi_\pm(z) = \exp \left\{ i \frac{\pi qM}{\im{\tau}} \im{\bar{\tau}z} \right\} \psi_\pm(z) ,
\end{eqnarray}
because of Eq.~(\ref{eq:A-bc}).
The magnetic flux should be quantized and $qM$ must be integer.

If $qM > 0$, $\psi_-$ has no zero-mode, but $\psi_+$ has $qM$ zero-modes and their 
wavefunctions are written as 
\begin{equation}
\psi^{j,M}_+(z) = \mathcal{N} e^{i\pi  qM z \frac{\im{z}}{\im{\tau}}} \cdot \vartheta \left[
\begin{array}{c}
\frac{j}{qM} \\
0
\end{array}
\right] \left( qM z, qM\tau \right),
\end{equation}
with $j=0,1,\cdots, (qM-1)$, 
where $\vartheta$ denotes the Jacobi theta function, 
\begin{equation}
\vartheta \left[
\begin{array}{c}
a \\
b
\end{array}
\right] (\nu, \tau) = \sum_{l \in {\bf Z}} e^{\pi i (a+l)^2 \tau} e^{2 \pi i (a+l)(\nu+b)} .
\end{equation}
Here, $\mathcal{N}$ denotes  the normalization factor given by 
\begin{equation}
\label{eq:normalization}
\mathcal{N} = \left( \frac{2\im{\tau} |qM|}{\mathcal{A}^2} \right)^{1/4}, 
\end{equation}
with $\mathcal{A}= 4 \pi^2 R^2 \im{\tau}$.
See Appendix \ref{sec:app-1} for computation of $\mathcal{N}$.

If $qM <0$, $\psi_+ $ has no zero-mode, but $\psi_-$ has $|qM|$ zero-modes.
Their wavefunctions are the same as the above except replacing $qM$ by $|qM|$.
Thus, introducing magnetic flux leads to a chiral theory.

For simplicity, we normalize the charge $q=1$.
We can discuss other charges $q \neq 1$ by replacing $M \rightarrow qM$ in the following 
analysis.
Hereafter, we also set $M >0$.
Thus, in what follows, we consider the zero-mode wavefunctions, 
\begin{equation}
\label{eq:basis1}
\psi^{j,M}(z,\tau) = \mathcal{N} \cdot e^{i \pi M z  \im{z}/ \im{\tau} } \cdot \vartheta \left[
\begin{array}{c}
\frac{j}{M} \\
0
\end{array}
\right] ( M z, M \tau) .
\end{equation}
Here, we write $\tau$ explicitly in $\psi^{j,M}(z,\tau)$ because $\tau$ dependence is important in the following analysis.
We can use another basis of zero-mode solutions, 
\begin{equation}
\label{eq:basis2}
\chi^{j,M}(\tau,z) = \frac{\mathcal{N}}{\sqrt{M}} \cdot e^{i \pi M z  \im{z}/ \im{\tau} } \cdot \vartheta \left[
\begin{array}{c}
0 \\
\frac{j}{M}
\end{array}
\right] ( z, \tau/M ).
\end{equation}
These are related with each other as 
\begin{eqnarray}
\label{eq:cjk}
\chi^{j,M} &=& \frac{1}{\sqrt{M}} \sum_k e^{2\pi i \frac{jk}{M}} \psi^{k,M}, \\
\label{eq:cjki}
\psi^{j,M} &=&  \frac{1}{\sqrt{M}} \sum_k e^{-2\pi i \frac{jk}{M}} \chi^{k,M} .
\end{eqnarray}
See Appendix \ref{sec:app-2} for these relations.

Using these wavefunctions, we can compute 3-point coupling \cite{Cremades:2004wa},
\begin{equation}
\label{eq:3-point}
\int d^2z \ \psi^{j_1,M_1}(z)\psi^{j_2,M_2}(z)\psi^{j_2,M_2}(z),
\end{equation}
as well as $n$-point couplings \cite{Abe:2009dr},
\begin{equation}
\label{eq:n-point}
\int d^2z \ \psi^{j_1,M_1}(z)\psi^{j_2,M_2}(z) \cdots \psi^{j_n,M_n}(z).
\end{equation}

\subsection{$T^2/Z_2$ orbifold}

In \cite{Abe:2008fi}, the zero-mode wavefunctions on the $T^2/Z_2$ orbifold were studied.
On the $T^2/Z_2$ orbifold, we identify $z \sim -z$.
Under the $Z_2$ twist, the zero-mode wavefunctions satisfy the following simple relation,
\begin{equation}
\label{eq:zto-z}
\psi^{j,M}(-z) = \psi^{M-j,M}(z). 
\end{equation}
Note that $\psi^{0,M}(z) = \psi^{M,M}(z)$.
The other basis, $\chi^{j,M}(z)$, also satisfies the same relation.
Thus, the $Z_2$ even and odd wavefunctions $\Theta^{j,M}_{\pm1}(z)$ can be written by 
\begin{eqnarray}
\label{eq:Z2-wf}
\Theta^{j,M}_{\pm1}(z) &=& \frac{1}{\sqrt{2}} \left( \psi^{j,M}(z) \pm \psi^{M-j,M}(z) \right).
\end{eqnarray}
The numbers of even and odd modes are shown in Table \ref{tab:Z2}.

\begin{table}[h]
\begin{tabular}{|c|c|c|} \hline
M & $2n$ & $2n+1$ \\ \hline \hline
$Z_2$ even & $n+1$ & $n+1$ \\  \hline 
$Z_2$ odd  & $n-1$ &  $n$   \\  \hline
\end{tabular}
\caption{The numbers of $Z_2$ even and odd zero-modes.}
\label{tab:Z2}
\end{table}

By using $Z_2$ eigenfunctions, $\Theta^{j,M}_{\pm1}(z)$, we can compute 
3-point couplings and higher order couplings similar to Eqs.(\ref{eq:3-point}) 
and (\ref{eq:n-point}).
Then, we obtain phenomenological interesting results 
e.g., realization of quark and lepton mass hierarchies and their mixing angles \cite{Abe:2012fj,Abe:2014vza,Kobayashi:2015siy,Kobayashi:2016qag}.

We also give a comment on Scherk-Schwarz phases and discrete Wilson lines.
These degrees of freedom are equivalent to each other \cite{Abe:2013bca}.
Hence, we restrict ourselves to Scherk-Schwarz phases.
With Scherk-Schwarz phases $(\beta_1,\beta_\tau)$, the boundary conditions (\ref{eq:z+1}) and  (\ref{eq:z+tau}) change as 
\begin{eqnarray}
\label{eq:z+1-SS}
\psi (z+1) &=& e^{i\phi_1(z) + 2\pi i\beta_1}\psi(z), \\
\label{eq:z+tau-SS}
\psi (z + \tau) &=& e^{i\phi_2(z) + 2\pi i \beta_\tau}\psi(z) ,
\end{eqnarray}
for $q=1$.
On the orbifold, discrete values of Scherk-Schwarz phases are possible \cite{Abe:2013bca}.
(See also \cite{Kobayashi:1990mi}.)
On the $T^2/Z_2$ orbifold, there are four possible Scherk-Schwarz phases, 
\begin{equation}
(\beta_1,\beta_\tau)=(0,0), \qquad (0,1/2), \qquad (1/2,0), \qquad (1/2,1/2).
\end{equation}
For such boundary conditions, the zero-mode wavefunctions are obtained as \cite{Abe:2013bca}
\begin{equation}
\label{eq:basis1-WL}
\psi^{j+\beta_1,\beta_\tau,M}(z ) = \mathcal{N} \cdot e^{i \pi M z  \im{z}/ \im{\tau} } \cdot \vartheta \left[
\begin{array}{c}
\frac{j+\beta_1}{M} \\
-\beta_\tau
\end{array}
\right] ( M z, M \tau) .
\end{equation}
Under the $Z_2$ twist, these wavefunctions behaves as 
\begin{equation}
\psi^{j+\beta_1,\beta_\tau,M}(-z) = \psi^{M-j-\beta_1,-\beta_\tau,M}(z )
=e^{-4\pi i \frac{(j+ \beta_1)\beta_\tau}{M}}\psi^{M-j-\beta_1,\beta_\tau,M}(z ) .
\end{equation}
Using this behavior, we can construct $Z_2$ eigenstates similar to Eq.(\ref{eq:Z2-wf}).

\section{$T^2/Z_4$ orbifold}
\label{sec:3}

Here, we study $T^2/Z_4$ orbifold models.

\subsection{Modular transformation}

We denote the basis vectors of the lattice $\Lambda$ by $(\alpha_1,\alpha_2)$ to construct 
$T^2 = R^2/\Lambda$, i.e., 
$\alpha_1 = 2\pi R$ and $\alpha_2 = 2\pi R \tau$ in the complex basis.
The same lattice can be described by another basis, $(\alpha'_1,\alpha'_2)$, 
and these lattice bases are related with each other as,
\begin{equation}
\label{eq:SL2Z}
\left(
\begin{array}{c}
\alpha'_2 \\ \alpha'_1
\end{array}
  \right) =\left(
  \begin{array}{cc}
a & b \\
c & d   
\end{array}
\right) \left(
  \begin{array}{c}
\alpha_2 \\ \alpha_1
\end{array}
  \right) ,
\end{equation}
where $a,b,c,d$ are integer with satisfying $ad-bd = 1$.
That is $SL(2,Z)$ transformation.
The lattice basis $(\alpha_1,\alpha_2)$ spans  exactly the same lattice as the basis $(-\alpha_1,-\alpha_2)$.
Thus, the modular transformation is $SL(2,Z)/Z_2$ .

Under the above transformation (\ref{eq:SL2Z}), the 
modular parameter $\tau$ transforms as 
\begin{equation}
\tau \longrightarrow  \frac{a\tau + b}{c \tau + d}.
\end{equation}
This transformation includes two important generators, $S$ and $T$,
\begin{eqnarray}
& &S:\tau \longrightarrow -\frac{1}{\tau}, \\
& &T:\tau \longrightarrow \tau + 1.
\end{eqnarray}
Here, we study $S$ because it is relevant to the $Z_4$ twist.
$S$ transforms the lattice basis as 
\begin{equation}
(\alpha_1,\alpha_2)  \longrightarrow (-\alpha_2,\alpha_1).
\end{equation}
This is nothing but  the $Z_4$ twist, for $\tau = i$.
More precisely we can refer to this as the inverse of the $Z_4$ twist, i.e., the $-\pi/2$ rotation.

\subsection{$T^2/Z_4$ orbifold model}

Here, we study the transformation behavior of zero-mode wavefunctions under $S$.
Let us start with $\chi^{j,M}(z,\tau)$.
Then, we examine its $S$ transformation.
That is, we replace $\tau \rightarrow -1/\tau$, $z \rightarrow z/\tau$ in $\chi^{j,M}(z,\tau)$.
It is found that 
\begin{equation}
\label{eq:Z4chi-psi}
\chi^{j,M}(z/\tau,-1/\tau) = \psi^{j,M}(z,\tau).
\end{equation}
To show this transformation, we have used the following relation,
\begin{equation}
\label{eq:theta-relation1}
\vartheta \left[
\begin{array}{c}
0 \\
a
\end{array}
\right] \left( \frac{\nu}{\kappa}, - \frac{1}{\kappa} \right) = (-i \kappa)^{1/2} e^{i \pi \nu^2/\kappa} \cdot \vartheta \left[
\begin{array}{c}
a \\
0
\end{array}
\right] \left( \nu, \kappa \right).
\end{equation}
That is, the 
$\vartheta$ function in $\chi^{j,M}(z,\tau)$ transforms 
\begin{equation}
\label{eq:Z4wf-S}
\vartheta \left[
\begin{array}{c}
0 \\
\frac{j}{M}
\end{array} \right] \left( z, \frac{\tau}{M} \right) \to \left( -i M \tau \right)^{1/2} e^{i \pi M \frac{z^2}{\tau}} \cdot \vartheta \left[
\begin{array}{c}
\frac{j}{M} \\
0 
\end{array} \right] \left( M z, M \tau \right) .
\end{equation}
In addition, we combine the $S$ transformation of the phase  $ e^{i \pi M z \frac{\im{z}}{\im{\tau}}}$ with 
the phase factor $e^{i \pi M \frac{z^2}{\tau}}$ in the above equation (\ref{eq:Z4wf-S})  to find 
\begin{eqnarray}
\exp \left\{ \pi i M \frac{z}{\tau} \frac{\im{z/\tau}}{\im{(-1/\tau)}}  + \pi i M \frac{z^2}{\tau} \right\} 
&=& \exp \left\{  \pi i M \frac{z \cdot \im{z}}{\im{\tau}} \right\} .
\end{eqnarray}
Also the normalization factor transforms under $S$,
\begin{equation}
\mathcal{N} \to \left(\frac{1}{|\tau|^2} \right)^{1/4} \mathcal{N}.
\end{equation}
Using these results, we can derive the transformation (\ref{eq:Z4chi-psi}) \cite{Cremades:2004wa}.\footnote{
Such a transformation behavior is important in modular symmetry of 4D low-energy effective field theory \cite{Kobayashi:2016ovu}.}

On the other hand,  we replace 
\begin{equation}
\label{eq:S-Z4}
\tau \rightarrow -1/\tau, \qquad z \rightarrow \tau z ,
\end{equation}
in $\psi^{j,M}(z,\tau)$.
Similarly, we find that 
\begin{equation}
\psi^{j,M} \left( \tau z, -\frac{1}{\tau} \right) = \chi^{j,M} (z ,\tau).
\end{equation}

We require that the torus is invariant under the $S$ transformation, i.e.
\begin{equation}
\tau = -\frac{1}{\tau}.
\end{equation}
Its solution is $\tau = \pm i$.
Here, we set $\tau = i$.
Then, the above transformation (\ref{eq:S-Z4})  is nothing but the  $Z_4$ twist, 
$z \rightarrow \tau z = iz$.
Thus, under such $Z_4$ twist, wavefunctions transform,
\begin{eqnarray}
\psi^{j,M}(z,\tau=i) \rightarrow & & \psi^{j,M}(iz,-1/\tau = i )   \nonumber \\
                             & &= \chi^{j,M}(z,\tau = i) \\ 
& &= C^j_{k,M}\psi^{k,M}(z,\tau=i).    \nonumber 
\end{eqnarray}
In the last equality, we have used the relation (\ref{eq:cjk}), and 
the coefficients $C^j_{k,M}$ are written by 
\begin{equation}
C^{j}_{k,M} = \frac{1}{\sqrt{M}} e^{2\pi i \frac{jk}{M}}  .
\end{equation}

The matrix $C^{j}_{k,M}$ satisfies
\begin{equation}
\sum_k C^{j}_{k,M}C^{k}_{l,M} = \frac{1}{M} \sum_k e^{2 \pi i (j+l)k/M} = \delta_{(j+l),~nM},
\end{equation}
where $n$ is integer.
That is, we find that the $Z_4$ transformation, 
\begin{equation}
\psi^{j,M}(z,i) \rightarrow \chi^{j,M}(z,i) \rightarrow \psi^{M-j,M}(z,i) \rightarrow \chi^{M-j,M}(z,i) \rightarrow \psi^{j,M}(z,i).
\end{equation}
This transformation property is consistent with the $Z_2$ transformation (\ref{eq:zto-z}).
That is, we can write 
\begin{equation}
\psi^{j,M}(z,i) \rightarrow \chi^{j,M}(z,i) \rightarrow \psi^{j,M}(-z,i) \rightarrow \chi^{j,M}(-z,i) \rightarrow \psi^{j,M}(z,i),
\end{equation}
and operation of the  $Z_4$ twist two times is just the $Z_2$ twist.

Now, we can write the zero-mode wavefunctions with $Z_4$ eigenvalues $\gamma =\pm 1, \pm i$ as 
\begin{equation}
\frac12 \left(   \psi^{j,M}(z,i) +\gamma^{-1} \chi^{j,M}(z,i) + \gamma^{-2} \psi^{M-j,M}(z,i)  + \gamma^{-3} \chi^{M-j,M}(z,i)  \right),
\end{equation}
i.e.,
\begin{equation}
\frac12 \left(   \psi^{j,M}(z,i) +\gamma^{-1} \sum_k C^j_{k,M}\psi^{k,M}(z,i) + \gamma^{-2} \psi^{M-j,M}(z,i)  + \gamma^{-3}  \sum_k C^{M-j}_{k,M}\psi^{k,M}(z,i)  \right).
\end{equation}

Obviously, we can construct the $Z_4$ eigenstates as those of the matrix $C^j_{k,M}$.
As an illustrating example, we study the model with $M=3$, where the matrix $C^j_{k,M}$ is 
obtained as 
\begin{equation}
C^{j}_{k,M} = \frac{1}{\sqrt{3}} \left(
\begin{array}{ccc}
1 & 1 & 1 \\
1 & \rho & \rho^2 \\
1 & \rho^2 & \rho
\end{array} \right) ,
\end{equation}
with $\rho ={2\pi i /3}$.\footnote{This matrix is the same as the matrix representation of $S$ in 
heterotic string theory on the $Z_3$ orbifold \cite{Lauer:1989ax}.}
This matrix has eigenvalues, $\gamma = 1, -1, i$, and 
eigenvectors in the basis $\sum_j a_j\psi^{j,3}$,
\begin{eqnarray}
(a_0,a_1,a_2) = & & (1 + \sqrt{3}, 1,1) {\rm ~~~~for~~~~} \gamma=1, \nonumber \\
& & (1 - \sqrt{3}, 1,1) {\rm ~~~~for~~~~} \gamma=-1, \\
& & (0, 1,-1) {\rm ~~~~for~~~~} \gamma=i,   \nonumber 
\end{eqnarray}
up to normalization factors.

Similarly, we can obtain $Z_4$ eigenvalues and eigenstates by using explicit matrices, $C^j_{k,M}$ 
for each value of $M$, in particular small values of $M$.
Table \ref{tab:Z4-eigen} shows the numbers of  $Z_4$ zero-modes for small values of $M$.
This result is consistent with the previous results \cite{Abe:2013bca,Abe:2014noa} 
up to the definition of the $Z_4$ twist.\footnote{
The number of zero-modes with $Z_4$ eigenvalue $\gamma = i$ is exchanged for 
the number of zero-modes with eigenvalue $\gamma = -i$ when we replace the definition 
of $Z_4$ twist by its inverse.  }
The corresponding $Z_4$ eigenstates are shown in Appendix \ref{sec:app-3}.

We give a comment on $Z_4$ eigenstates.
The $Z_2$ even states, $(\psi^{j,M}+\psi^{M-j,M})$, correspond to 
the $Z_4$ eigenstates with eigenvalues $\gamma = \pm 1$, 
while $Z_2$ odd states, $(\psi^{j,M}-\psi^{M-j,M})$, correspond to 
the $Z_4$ states with eigenvalues $\gamma = \pm i$.
Explicit results on eigenstates for small number of $M$ are shown in 
Appendix \ref{sec:app-3}.
For $M=$ even, $Z_4$ eigenvectors are relatively simple, while 
for $M=$ odd $Z_4$ eigenvectors are complicated.

\begin{table}[h]
\begin{tabular}{|c||c|c|c|c|c|c|c|c|c|c|c|c|} \hline
$M$ & 1 & 2 & 3 & 4 & 5 & 6 & 7 & 8 & 9 & 10 & 11 & 12 \\ \hline \hline
$Z_4$ eigenvalue : $+1$ &1&1&1&2&2&2&2&3&3&3&3&4 \\ \hline
$Z_4$ eigenvalue : $-1$ &0&1&1&1&1&2&2&2&2&3&3&3 \\ \hline
$Z_4$ eigenvalue : $+i$ &0&0&1&1&1&1&2&2&2&2&3&3 \\ \hline
$Z_4$ eigenvalue : $-i$ &0&0&0&0&1&1&1&1&2&2&2&2 \\ \hline
\end{tabular}
\caption{The number of zero-modes in the $Z_4$ orbifold model.}
\label{tab:Z4-eigen}
\end{table}

From the above explicit results, we can expect generic results on the numbers of 
zero-modes, which are shown in Table \ref{tab:Z4-generic}.
Indeed,  we can prove this result.
First, we compute ${\rm tr}\ C^{j}_{k,M}$, 
\begin{eqnarray}
\label{eq:trCe}
\mathrm{tr}\  C &=& \frac{1}{\sqrt{M}} \sum_{k=0}^{M-1} e^{2 \pi i \frac{k^2}{M}} .
\end{eqnarray}
In our computation, the following Landsberg-Schaar relation:  
\begin{equation}
\label{eq:LSR}
\frac{1}{\sqrt{p}} \sum^{p-1}_{n=0} e^{\frac{2\pi i n^2 q}{p}} = \frac{e^{\frac{\pi i }{4}}}{\sqrt{2q}} \sum^{2q-1}_{n=0} e^{-\frac{\pi i n^2 p}{2q}},\ \ p,q \in \mathbb{N} ,
\end{equation}
is very useful.
We take $p=M$, $q=1$ in the Landsberg-Schaar relation to compute ${\rm tr}\ C$,
\begin{eqnarray}
{\rm tr} \ C &=& e^{\frac{\pi i}{4} } \frac{1}{\sqrt{2}} \sum^1_{k=0} e^{- \frac{\pi i k^2 M}{2}} 
= e^{\frac{\pi i}{4} } \frac{1}{\sqrt{2}} \left( 1 + e^{-\pi i \frac{M}{2}} \right).
\end{eqnarray}
Then, we find that 
\begin{equation}
\label{eq:trC}
{\rm tr} \ C = \left\{ 
\begin{array}{cl}
1+i & {\rm ~~~for} \ M=4n \\
1 &{\rm ~~~for} \ M=4n+1 \\
0 &{\rm ~~~for} \ M=4n+2 \\
i &{\rm ~~~for} \ M=4n+3 
\end{array}
\right.   .
\end{equation}

\begin{table}[h]
\centering
\begin{tabular}{|c||c|c|c|c|} \hline
$M$   & $4n$ & $4n+1$ & $4n+2$ & $4n+3$ \\ \hline \hline
$Z_4$ eigenvalue : $+1$ & $n+1$ & $n+1$ & $n+1$ & $n+1$ \\ \hline
$Z_4$ eigenvalue : $-1$ & $n$ & $n$ & $n+1$ & $n+1$  \\ \hline
$Z_4$ eigenvalue : $+i$ & $n$ & $n$ & $n$ & $n+1$  \\ \hline
$Z_4$ eigenvalue : $-i$ & $n-1$ & $n$ & $n$ & $n$  \\ \hline
\end{tabular}
\caption{Generic results on the numbers of $Z_4$ zero-modes.}
\label{tab:Z4-generic}
\end{table}

For example, recall that when $M=4n$, there are $(2n+1)$ $Z_2$ even zero-modes and 
$(2n-1)$ $Z_2$ odd zero-modes.
That is, the sum of the numbers of $Z_4$ zero-modes with eigenvalues $\gamma = \pm 1$ is 
equal to $(2n+1)$, while the sum of  the numbers of $Z_4$ zero-modes with eigenvalues $\gamma = \pm i$ is equal to 
$(2n-1)$ .
Combination of these with Eq.(\ref{eq:trC}) leads to the result for $M=4n$ in Table \ref{tab:Z4-generic}.
Similarly, we can derive the numbers of $Z_4$ zero-modes with other values of $M$ 
as shown in Table \ref{tab:Z4-generic}.


\section{$T^2/Z_3$ orbifold}
\label{sec:4}

In this section, we study the zero-modes on 
$T^2/Z_3$ and $T^2/Z_6$ orbifolds.

\subsection{$T^2/Z_3$ orbifold}

Here, we study the $Z_3$ orbifold models.
Our strategy is the same as one in the previous section.
That is, we examine the modular transformation corresponding to the 
$Z_3$ twist.
A good candidate for the $Z_3$ twist is $ST$ transformation, 
because it satisfy $(ST)^3=1$ on $\tau$.
Under $ST$, the modular parameter $\tau$ transforms as 
\begin{equation}
\label{eq:tau-ST3}
\tau \to - \frac{1}{\tau+1}.
\end{equation}
When $\tau = e^{\pm 2\pi i /3}$, the modular parameter is invariant under $ST$, i.e.
\begin{equation}
\label{eq:tau-ST}
\tau = - \frac{1}{\tau+1}.
\end{equation}
For such a transformation, the $Z_3$ twist (its inverse) can be defined by 
 \begin{equation}
z \to \tau  z ,
\end{equation}
when  $\tau = e^{ 2\pi i /3}$ $(\tau = e^{- 2\pi i /3})$.
Alternatively, we can define the $Z_3$ twist by 
\begin{equation}
\label{eq:z-ST3}
z \to \frac{-z}{\tau+1},
\end{equation}
because of the relation (\ref{eq:tau-ST}).
In what follows, we study the transformation of wavefunctions under 
Eqs.~(\ref{eq:tau-ST3}) and (\ref{eq:z-ST3}).
We restrict ourselves to the models with $M=$ even, because 
the following transformation behavior is valid only for $M=$ even.

We find that 
\begin{eqnarray}
\chi^{j,M}(-z/(\tau + 1),-1/(\tau +1)) &=& e^{\pi i \frac{j^2}{M}} \cdot \psi^{j,M}(-z, \tau) \nonumber \\
&=&e^{\pi i \frac{j^2}{M}} \cdot \psi^{M-j,M}(z, \tau).
\end{eqnarray}
Here, we have used  the relation (\ref{eq:theta-relation1}) and the following relation, 
\begin{equation}
\label{eq:tau+1}
\theta \left[
\begin{array}{c}
a \\
b
\end{array} \right] (\nu,\tau+1) = e^{-i \pi a(a-1)} \cdot \theta \left[
\begin{array}{c}
a \\
b +a - \frac{1}{2} 
\end{array} \right] (\nu,\tau) .
\end{equation}
Since $\psi^{M-j,M} = C_{jk} \chi^{k,M}$, the transformation in the $\chi$ basis is written by 
\begin{equation}
\chi^{j,M}(z,\tau) \rightarrow D^{j}_{k,M}\chi^{k,M}(z,\tau), \qquad 
D^{j}_{k,M} = e^{\pi i \frac{j^2}{M}} C^j_{k,M}.
\end{equation}

When we examine the inverse transformation, 
\begin{equation}
\tau \to -\frac{1}{\tau} -1, 
\qquad 
z \to \frac{1}{\tau}z  ,
\end{equation}
on the wavefunction $\psi^{j,M}(z,\tau)$, 
we find that 
\begin{eqnarray}
\label{eq:z3psitochi}
\psi^{j,M} \left( \frac{z}{\tau}, -1/\tau -1  \right) &=& e^{-\pi i \frac{j^2}{M}} \cdot \chi^{M-j}(z, \tau).
\end{eqnarray}
Thus, it is found that under the above inverse transformation, 
the wavefunction $\psi^{j,M}$ transforms as 
\begin{equation}
\psi^{j,M} \to (D^{-1})^j_{k,M}\psi^{j,M},
\end{equation}
where $D^{-1}$ is the inverse matrix of $D^j_{k,M}$.

For example, for $M=2$, we obtain 
\begin{equation}
D^j_{k,M=2} = \frac{1}{\sqrt{2}} \left( 
\begin{array}{cc}
1 & 1 \\
i & -i
\end{array} \right) .
\end{equation}
However, we find that 
\begin{equation}
(D^j_{k,M=2})^3 = \frac{1}{\sqrt{2}} \left( 
\begin{array}{cc}
1+i & 0 \\
0 & 1+i
\end{array} \right) .
\end{equation}
This matrix does not realize exactly the $Z_3$ twist.

Indeed, for generic even number $M$, we can find 
\begin{equation}
D^j_{k,M} D^k_{\ell,M} D^\ell_{m,M} =\frac{1}{\sqrt{2}}(1+i)\delta_{j,m} = e^{\pi i /4} \delta_{j,m}.
\end{equation}
See Appendix \ref{sec:app-2}.
Thus, the matrix $D^{j}_{k,M}$ on $\chi^{j,M}$ does not represent the $Z_3$ twist exactly.

Here, we allow the constant phase for all modes under the above transformation, e.g. \footnote{
We have other two values for candidates of the constant phase, 
and totally there are three possibilities.
Different constant phases lead to change of degeneracy factors for each $Z_3$ eigenvalues.
Such possibilities of constant phases may correspond to the possibility of 
introduction of Scherk-Schwarz phases.
Similarly, we have the degree of freedom to define the $Z_4$ twist by $e^{\pi in/2}C^j_{k,M}$ with $n=0,1,2,3$.}
\begin{equation}
\chi^{j,M}(z,\tau) \rightarrow \tilde D^{j}_{k,M}\chi^{k,M}(z,\tau), \qquad 
\tilde D^{j}_{k,M} = e^{-\frac{\pi i}{12} }  D^{j}_{k,M}  .
\end{equation}
Then, we can realize the $Z_3$ twist, 
\begin{equation}
\tilde D^j_{k,M} \tilde D^k_{\ell,M} \tilde D^\ell_{m,M} =  \delta_{j,m}.
\end{equation}
Here, we employ this matrix $\tilde D$ as the $Z_3$ twist.

For example, for $M=2$, we use the following matrix for the $Z_3$ twist 
on $\chi^{j,M}$
\begin{equation}
\tilde D^j_{k,M=2} = \frac{ e^{-\frac{\pi i}{12} }}{\sqrt{2}} \left( 
\begin{array}{cc}
1 & 1 \\
i & -i
\end{array} \right) .
\end{equation}
Its eigenvalues are obtained as $\gamma = 1, e^{-2 \pi i/3}$, and 
eigenvectors are given as 
\begin{equation}
(1, \sqrt2  \gamma e^{\frac{\pi i}{12} }-1),
\end{equation}
in the $(\chi^{0,M},\chi^{1,M})$ basis, up to normalization factor.

Similarly, we study the model with $M=4$.
The eigenvalues of the matrix $\tilde D^j_{k,M}$ are 
\begin{equation}
(1,e^{2\pi i/3},e^{2\pi i/3},e^{-2 \pi i/3}),
\end{equation}
and their eigenvectors are obtained in the basis $a_i\chi_i$,
\begin{eqnarray}
& & \left[ 0,-1,0,1\right],  \nonumber \\
& & 
\left[ -i(-1+\sqrt{2}),(1+i)-\sqrt{2},1,0 \right] , \\
& & 
\left[ \frac{(-6+6i)+3i\sqrt{2}-(2+2i)\sqrt{3}+(1-2i)\sqrt{6}}{3i\sqrt{2}+(2-2i)\sqrt{3}},1,\frac{\sqrt{2}(3i +(1+2i)\sqrt{3})}{3i\sqrt{2}+(2-2i)\sqrt{3}+\sqrt{6}},1 \right] ,
\nonumber \\
& & 
\left[ \frac{6i +3(-1)^{1/4} +(1+3i)\sqrt{3/2} +2\sqrt{3}}{3(-1)^{1/4} -(1-i)\sqrt{3/2} +2i\sqrt{3}},1, \frac{\sqrt{2}(-3i+(1+2i)\sqrt{3}}{-3i\sqrt{2}+(2-2i)\sqrt{3}+\sqrt{6}},1 \right] ,
\nonumber 
\end{eqnarray}
up to normalization factors.

Similarly, we can analyze the eigenvalues and eigenvectors for other $M$.
Table \ref{tab:Z3} shows the numbers of $Z_3$ zero-modes with each eigenvalue 
for small values of $M$.
This result is consistent with the previous results \cite{Abe:2013bca,Abe:2014noa}.
We can derive eigenvectors, but their explicit forms are, in general, very complicated.

\begin{table}[h]
\centering
\begin{tabular}{|l||c|c|c|c|c|c|} \hline
\multicolumn{1}{|c||}{$M$}&2&4&6&8&10&12 \\ \hline \hline
$Z_3$ eigenvalue : $1$ &1&1&3&3&3&5 \\ \hline
$Z_3$ eigenvalue : $e^{2 \pi i/3}$ &0&2&2&2&4&4 \\ \hline
$Z_3$ eigenvalue : $e^{-2 \pi i/3}$&1&1&1&3&3&3 \\ \hline
\end{tabular}
\caption{The number of zero-modes  in the $Z_3$ orbifold model.}
\label{tab:Z3}
\end{table}

We can analyze the number of $Z_3$ zero-modes for generic even number $M$.
First we compute the trace of the inverse of $\tilde D$,
\begin{equation}
(\tilde D^{-1})^j_k =  e^{\pi i \frac{1}{12}} \cdot e^{-\pi i \frac{j^2}{M}} \cdot C_{jk}^\dagger .
\end{equation}
That is, its trace is written by 
\begin{equation}
{\rm tr}(\tilde D^{-1}) =  e^{\pi i \frac{1}{12}} \sum_{k=0}^{M-1} 
e^{-\pi i \frac{k^2}{M}} \cdot e^{-2\pi i \frac{k^2}{M}} = 
 e^{\pi i \frac{1}{12}} \sum_{k=0}^{M-1} e^{-3\pi i \frac{k^2}{M}} .
\end{equation}
Here, we use the Landsberg-Schaar relation (\ref{eq:LSR}) with $p=3$ 
and $2q=M$.
Then, we find 
\begin{equation}
{\rm tr}(\tilde D^{-1}) =  \frac{i e^{-\pi i \frac{2}{3}}}{\sqrt 3}
\left( 1 + 2 e^{\pi i \frac{M}{3}}\right) .  
\end{equation}
Explicitly, we obtain the following results,
\begin{equation}
\label{eq:tr-D}
{\rm tr} \tilde D^{-1}  = \left\{ 
\begin{tabular}{cl}
$1+\omega$ & for $M=6n+2$ \\
$\omega^2$ & for $M=6n+4$ \\
$2 + \omega^2$ & for $M=6n$ 
\end{tabular}
\right.   ,
\end{equation}
where $\omega =e^{2\pi i/3}$.
Then, the trace of its inverse can be obtained by replacing 
$\omega \rightarrow \omega^2$, 
\begin{equation}
\label{eq:trC33}
{\rm tr} \tilde D  = \left\{ 
\begin{tabular}{cl}
$1+\omega^2$ & for $M=6n+2$ \\
$\omega$ & for $M=6n+4$ \\
$2 + \omega$ & for $M=6n$ 
\end{tabular}
\right.   .
\end{equation}
From this result, we can derive the number of $Z_3$ eigenstates as 
shown in Table \ref{tab:Z3-generic}.
Note that $1 + \omega + \omega^2 = 0$.

\begin{table}[h]
\centering

\begin{tabular}{|l||c|c|c|} \hline
\multicolumn{1}{|c||}{$M$}&$6n$&$6n+2$&$6n+4$ \\ \hline \hline
$Z_3$ eigenvalue : $1$ &$2n+1$&$2n+1$&$2n+1$ \\ \hline
$Z_3$ eigenvalue : $e^{2\pi i /3}$ &$2n$&$2n$&$2n+2$ \\ \hline
$Z_3$ eigenvalue : $e^{2 \pi i/3}$ &$2n-1$&$2n+1$&$2n+1$ \\ \hline
\end{tabular}
\caption{Generic results on $Z_3$ zero-modes. }
\label{tab:Z3-generic}
\end{table}

\subsection{$Z_6$ orbifold}

Obviously, the $Z_6$ twist can realized by the product of the $Z_2$ and 
$Z_3$ twists.
Also, recall that the $Z_2$ twist on $\psi^{j,M}(z)$ and $\chi^{j,M}(z)$ is realized by 
\begin{equation}
\psi^{j,M}(z) \rightarrow  \psi^{j,M}(-z)=  \psi^{M-j,M}(z), \qquad  
\chi^{j,M}(z) \rightarrow  \chi^{j,M}(-z)=  \chi^{M-j,M}(z).
\end{equation}
Here, we restrict ourselves to the models with $M=$ even.
From the analysis on the $T^2/Z_3$ orbifold, the $Z_6$ twist can be realized by 
\begin{equation}
F^j_{k,M} =  e^{\frac{\pi i}{12} }e^{-\pi i \frac{j^2}{M}} {C^j_{k,M}} .
\end{equation}
Again, using the Landsberg-Schaar relation (\ref{eq:LSR}), 
we compute the trace of $F^j_{k,M}$ matrix,
\begin{equation}
\label{eq:tr-F}
{\rm tr }\ F = e^{\frac{\pi i}{12} } \sum_k e^{\pi i \frac{k^2}{M}} = e^{\frac{\pi i }{12}} e^{\frac{\pi i}{4}}.
\end{equation}

The possible eigenvalues of $F$-matrix are $\gamma = \rho^{k}$ with $k=0,1,\cdots, 5$ and 
$\rho = e^{\pi i /3}$.
Here, we denote the number of zero-modes with eigenvalues $\gamma$ by $N_\gamma$.
Since $(F^j_{k,M})^3$ corresponds to the $Z_2$ twist, 
the zero-mode numbers, $N_\gamma$ must satisfy 
\begin{equation}
N_1 + N_{\rho^2} + N_{\rho^4} = n+1, \qquad 
N_\rho+ N_{\rho^3} + M_{\rho^5} = n-1,
\end{equation}
for $M=2n$.
Similarly, $(F^j_{k,M})^2$ corresponds to the $Z_3$ twist,
the zero-mode numbers must satisfy 
\begin{equation}
N_1 + N_{\rho^3} = 2n+1 ,
\end{equation} 
for $M=6n, 6n+2, 6n+4$,
\begin{equation}
N_\rho + N_{\rho^4} = \left\{
\begin{tabular}{cl}
$2n$ & for $M=6n, 6n+2$ \\
$2n+2$ & for $M=6n+4$
\end{tabular}
\right.     ,
\end{equation}
and
\begin{equation}
N_{\rho^2} + N_{\rho^5} = \left\{
\begin{tabular}{cl}
$2n-1$ & for $M=6n $ \\
$2n+1$ & for $M=6n+2, 6n+4$
\end{tabular}
\right.     .
\end{equation}
Combining these relations with the trace (\ref{eq:tr-F}), we find the number of eigenstates, which is shown in Table \ref{tab:Z6}.

\begin{table}[h]
\centering
\begin{tabular}{|l||c|c|c|} \hline
\multicolumn{1}{|c||}{$M$}&$6n$&$6n+2$&$6n+4$ \\ \hline \hline
eigenvalue : $1$ &$n+1$&$n+1$&$n+1$ \\ \hline
eigenvalue : $e^{\pi i/3}$ &$n$&$n$&$n+1$ \\ \hline
eigenvalue : $e^{2\pi i/3}$ &$n$&$n+1$&$n+1$ \\ \hline
eigenvalue : $e^{3\pi i/3}$ &$n$&$n$&$n$ \\ \hline
eigenvalue : $e^{4\pi i/3}$ &$n$&$n$&$n+1$ \\ \hline
eigenvalue : $e^{5\pi i/3}$ &$n-1$&$n$&$n$ \\ \hline
\end{tabular}
\caption{Generic results on $Z_6$ zero-modes. }
\label{tab:Z6}
\end{table}

In principle, we can derive zero-mode wavefunctions with eigenvalues $\gamma$, 
but its explicit form is complicated.

\section{Zero-modes on orbifolds}
\label{sec:5}

We have studied the zero-modes on several orbifolds, $T^2/Z_N$ with $N=2,3,4,6$.
Now, let us compare our results between different $T^2/Z_N$ orbifolds.
We examine the $Z_N$ invariant zero-modes.
It is found that the number of $Z_N$ invariant zero-modes is written by 
\begin{equation}
I_{M,N} = \lfloor M/N \rfloor + 1,
\end{equation}
on $T^2/Z_N$ orbifold with magnetic flux $M$ except the $Z_3$ orbifold with $M=6n+4$.
Here, $\lfloor r \rfloor$ denotes the maximum integer $n$, which satisfies 
$n \leq r$.
Alternatively, the number of $Z_3$ invariant zero-modes is written by 
\begin{equation}
I_{M,N}^{(3)} = 2\lfloor M/(2N) \rfloor + 1,
\end{equation}
Our results are quite universal for different $T^2/Z_N$ orbifolds.

The index theorem tells that the number of zero-modes of the Dirac operators 
on $T^2$ with flux $M$ is equal to $M$.
The above number $I_{M,N}$ as well as $I^{(3)}_{M,N}$ would correspond to such an index on the $T^2/Z_N$ orbifolds.


It would be useful to rewrite the numbers of zero-modes with other eigenvalues 
by using the symbol $\lfloor r \rfloor$.
These are shown in Table \ref{tab:ZN}.
Note that the number of zero-modes with $Z_N$ eigenvalue $\gamma$ is exchanged for 
one with $Z_N$ eigenvalue $\gamma^{-1}$ when we replace the definition of $Z_N$ twist by 
its inverse.

\begin{table}[h]
\centering
\begin{tabular}{|l||c|} \hline
 eigenvalues ($\gamma$)  & number of zero-modes \\ \hline \hline
$Z_{N=2n}$ invariant & $\lfloor M/N \rfloor + 1$ \\ \hline 
$Z_3$ invariant & $2\lfloor M/(2N) \rfloor + 1$ \\ \hline \hline
$Z_2$ ($\gamma = -1$)      & $\lfloor (M-1)/2 \rfloor $ \\ \hline \hline
$Z_4$ ($\gamma =-1$)       & $\lfloor (M-2)/4  \rfloor +1 $ \\ \hline 
$Z_4$ ($\gamma =i$)       & $\lfloor (M-3)/4  \rfloor +1 $ \\ \hline 
$Z_4$ ($\gamma =-i$)       & $\lfloor (M-1)/4 \rfloor $ \\ \hline \hline
$Z_3$ ($\gamma = e^{2 \pi i/3}$)       & $2\lfloor (M-4)/6  \rfloor +2$ \\ \hline 
$Z_3$ ($\gamma = e^{-2 \pi i/3}$)       & $2\lfloor (M-2)/6  \rfloor +1 $ \\ \hline \hline
$Z_6$ ($\gamma = e^{ \pi i/3}$)       & $\lfloor (M-4)/6  \rfloor  +1 $ \\ \hline 
$Z_6$ ($\gamma = e^{ 2\pi i/3}$)       & $\lfloor (M-2)/6 \rfloor +1$ \\ \hline 
$Z_6$ ($\gamma = e^{ 3\pi i/3}$)       & $\lfloor M/6  \rfloor $ \\ \hline 
$Z_6$ ($\gamma = e^{ 4\pi i/3}$)       & $\lfloor (M-4)/6  \rfloor +1 $ \\ \hline 
$Z_6$ ($\gamma = e^{ 5\pi i/3}$)       & $\lfloor (M-2)/6  \rfloor $ \\ \hline 
\end{tabular}
\caption{Generic results on $Z_N$ zero-modes. }
\label{tab:ZN}
\end{table}

It seems that the $T^2/Z_3$ orbifold has the zero-mode structure different from the other orbifolds.
The numbers of zero-modes on $T^2/Z_N$ with $N=$ even have the structure with the 
period $N$ for $M$.
That is, the number of zero-modes increases by one when we replace $M$ by $M+N$.
On the other hand, the number of zero-modes on $T^2/Z_3$ has the structure 
with the period 6, and the number of zero-modes increases by two 
when replace $M$ by $M+6$.
Such a structure of $T^2/Z_3$ is similar to one of $T^2/Z_6$, and seems to be originated 
from the $T^2/Z_6$ orbifold.
At any rate, the deep reason why the $T^2/Z_3$ orbifold has a different structure 
is not clear.
It is important to study its reason further.

The number of $Z_N$ invariant zero-modes depends on non-trivial Scherk-Schwarz phases 
and discrete Wilson lines.
Thus, our results imply that the number of $Z_N$ invariant zero-modes is 
universal over all of $T^2/Z_N$ orbifolds if we choose proper conditions 
on Scherk-Schwarz phases and discrete Wilson lines.


\section{Conclusion}
\label{sec:conclusion}

We have studied $T^2/Z_N$ orbifold models with magnetic flux.
We used the modular transformation to define the orbifolds.
Then, we have computed zero-mode wavefunctions with 
each eigenvalue of the $Z_N$ twist.
We have shown the zero-mode numbers.
It is found that the number of the $Z_N$ invariant zero-modes 
is universal among different $T^2/Z_N$ orbifolds, 
and it can be obtained by  $\lfloor M/N \rfloor +1$ except one case in the $T^2/Z_3$ orbifold.
The zero-mode number of the Dirac operator on $T^2$ is given by 
$M$.
Our  result would correspond to such an index.

We can write wavefunctions analytically for fixed $M$.
Thus, we can compute 3-point couplings and higher order couplings.
Hence, our results would be useful to further phenomenological applications.
One can also apply our method to not only zero-modes, but also higher modes.


\section*{Acknowledgments}
T.~K. was is supported in part by the Grant-in-Aid for Scientific Research 
 No.~26247042  and No.~17H05395 from the Ministry of Education, Culture, Sports,
 Science and Technology in Japan.

%


\appendix

\section{Normalization of wavefunction and inner product of 
$\psi$ and $\chi$}
\label{sec:app-1}

In this appendix, we show computations on normalization  $\mathcal{N}$ of wavefunctions 
and the relations (\ref{eq:cjk}) and  (\ref{eq:cjki}).
The computation on normalization is useful for computation of 
the relations (\ref{eq:cjk}) and  (\ref{eq:cjki}).
Now, we compute 
\begin{equation}
\label{eq:norm}
\int_{T^2} dz d \bar{z} \psi^j(\psi^k)^* ,
\end{equation}
where 
\begin{equation}
\int_{T^2} dz d\bar{z} = 
 \mathcal{A} \int^1_0 d(\re{z}) \int^1_0 d \left(\frac{\im{z}}{\im{\tau}} \right),
\end{equation}
with $\mathcal{A} = 4 \pi^2 R^2 \im{\tau}$.
The product of the wavefunctions, $ \psi^j(\psi^k)^*$ is written 
explicitly, 
\begin{eqnarray}
\psi^j_{\pm}(\psi^k_{\pm})^* &=& \psi^{j,qM}(\tau,z) \cdot \psi^{-k,-qM}(\bar{\tau},\bar{z}) \\
&=& \mathcal{N}^2 \cdot e^{-2\pi q M (\im{z})^2/\im{\tau}} \cdot \vartheta \left[
\begin{array}{c}
\frac{j}{qM} \\
0
\end{array}
\right] \left( qMz, qM\tau \right) \cdot \vartheta \left[
\begin{array}{c}
\frac{k}{qM} \\
0
\end{array}
\right] \left( -qM\bar{z}, -qM \bar{\tau} \right). \nonumber 
\end{eqnarray}
The product of theta functions includes the following terms depending on 
$\re{z}$ and $\im{z}$, 
\begin{equation}
\sum_n \sum_{n'} e^{2 \pi i \left\{ \left( \frac{j}{qM} +n \right) - \left( \frac{k}{qM} +n'\right) \right\} \re{z}} \cdot e^{-2 \pi \left\{ \left( \frac{j}{qM} +n \right) + \left( \frac{k}{qM} +n'\right) \right\} \im{z} } .
\end{equation}
Then, the integration over $\re{z}$ leads to the Kronecker delta, 
$
\delta_{ {j}/{(qM)} +n , ~{k}/{(qM)} +n' } .
$
Thus, we obtain 
\begin{equation}
\int^1_0 d(\re{z})  \psi^{j,qM}(\tau,z) \cdot \psi^{-k,-qM}(\bar{\tau},\bar{z}) 
= \mathcal{N}^2 \sum_n e^{-2\pi qM \im{\tau} \left( n +\frac{j}{qM} + \frac{\im{z}}{\im{\tau}} \right)^2} .
\end{equation}
Furthermore, we can find 
\begin{eqnarray}
\label{eq:int-im}
\int^1_0 d\left( \frac{\im{z}}{\im{\tau}} \right) \sum_n e^{-2\pi qM \im{\tau} \left( n +\frac{j}{qM} + \frac{\im{z}}{\im{\tau}} \right)^2} &=& \sum_n \int^1_0 d\left( \frac{\im{z}}{\im{\tau}} \right)  e^{-2\pi qM \im{\tau} \left( n +\frac{j}{qM} + \frac{\im{z}}{\im{\tau}} \right)^2} \nonumber  \\
&=& \int^{\infty}_{\infty} dx e^{-2\pi qM \im{\tau} x^2} \\
&=& \left( \frac{1}{2 qM \im{\tau}} \right)^{\frac{1}{2}} . \nonumber 
\end{eqnarray}
Then, we find the normalization factor (\ref{eq:normalization}).

Similarly, we compute
\begin{equation}
\int dz d\bar{z} \  \chi^{j,M}(z, \tau) \cdot \left( \psi^{k,M} (z,\tau) \right)^*,
\end{equation}
where 
\begin{equation}
\chi^{j,M} \cdot \psi^{-k,-M} = \frac{(\mathcal{N})^2}{\sqrt{M}} \cdot e^{-2 \pi M \frac{(\im{z})^2}{\im{\tau}}} \cdot \theta \left[
\begin{array}{c}
0 \\
\frac{j}{M}
\end{array} \right] \left( z, \frac{\tau}{M} \right) \cdot \theta \left[
\begin{array}{c}
\frac{k}{M} \\
0 
\end{array} \right] (-\bar{z} M, -\bar{\tau} M)  .
\end{equation}

The product of the theta functions includes the following terms 
depending on $\re{z}$,
\begin{equation}
\sum_l \sum_{l'} e^{2 \pi i \left\{ l - M \left( l' + \frac{k}{M} \right) \right\} \re{z}}  .
\end{equation}
The integration over  $\re{z}$ leads to the Kronecker delta, 
$\delta_{\ell, ~M \ell' + k}$ .
Thus, we obtain 
\begin{equation}
\int^1_0 \re(z) \chi^{j,M} \cdot \psi^{-k,-M} = 
e^{2 \pi i \frac{jk}{M}} \cdot \mathcal{N}^2 \cdot e^{-2 \pi M \frac{(\im{z})^2}{\im{\tau}}} \sum_l e^{-2 \pi M \im{\tau} \left( l +\frac{k}{M} \right)} \cdot e^{-4 \pi M \left( l +\frac{k}{M} \right) \im{z}} .
\end{equation}
In addition, we can integrate this over $\im(z)$ similar to 
Eq.(\ref{eq:int-im}).
Then, we can derive 
\begin{equation}
\label{eq:psichi}
\int dz d\bar{z} \ \chi^{j,M}(z, \tau) \cdot \left( \psi^{k,M} (z,\tau) \right)^* = e^{2 \pi i \frac{jk}{M}} .
\end{equation}
That is nothing but the relation (\ref{eq:cjk}).

Also we can obtain the complex conjugate of Eq.(\ref{eq:psichi}),
\begin{equation}
\int dz d\bar{z} \ \psi^{j,M}(z, \tau) \cdot \left( \chi^{k,M} (z,\tau) \right)^* = e^{-2 \pi i \frac{jk}{M}},
\end{equation}
and this is nothing but the relation (\ref{eq:cjki}).

\section{Computation of $(D)^3$}
\label{sec:app-2}

In this section, we give the computation on $(D^j_{k,M})^3$ for generic even number $M$.
First, we can obtain
\begin{eqnarray}
D^j_{k,M} D^k_{\ell,M} &=&  \frac{1}{M}\sum_k e^{\frac{2 \pi i }{M}\left( 
\frac{j^2}{2} + k(j +\ell ) + \frac{k^2}{2}\right) } \nonumber \\
&=&  \frac{1}{M}\sum_k e^{\frac{ \pi i }{M} \left[  (k+j+ \ell)^2 - 
\ell (2j + \ell)  
\right] } \\
& =& \frac{1}{\sqrt{2M}} (1+i) e^{\frac{ \pi i }{M} \left[ -\ell(2j+\ell) \right]}  .
\nonumber
\end{eqnarray}
We have used the Landsberg-Schaar relation (\ref{eq:LSR}).
Then, we can compute 
\begin{eqnarray}
D^j_{k,M} D^k_{\ell,M} D^\ell_{m,M} &=&  \frac{1}{\sqrt 2 M}(1+i) \sum_\ell 
e^{\frac{ \pi i }{M} \left[ -\ell(2j+\ell) + \ell^2 + 2\ell m \right]}  \nonumber \\
&=& 
 \frac{1}{\sqrt 2 M}(1+i) \sum_\ell e^{\frac{ 2\pi i }{M}\ell (m-j)} \\
& = & \frac{1}{\sqrt 2}(1 + i) \delta_{j,m} = e^{\pi i /4} \delta_{j,m}. \nonumber
\end{eqnarray}
Again, we have used the Landsberg-Schaar relation (\ref{eq:LSR}).

\section{Eigenvectors in $Z_4$ orbifold models}
\label{sec:app-3}

In this section, we give explicitly 
$Z_4$ eigenvectors for $M=2,\cdots, 12$.
These eigenvectors are represented in the basis 
$\sum_{k = 0}^{M-1} a_k \psi^{k,M}$.
The $Z_2$ even states, $(\psi^{j,M}+\psi^{M-j,M})$, correspond to 
the $Z_4$ eigenstates with eigenvalues $\gamma = \pm 1$, 
while $Z_2$ odd states, $(\psi^{j,M}-\psi^{M-j,M})$, correspond to 
the $Z_4$ states with eigenvalues $\gamma = \pm i$.
Thus, $\psi^{0,M}$ does not correspond to eigenstates with $Z_4$ eigenvalues $\gamma = \pm i$, 
but always appears as eigenstates with $Z_4$ eigenvalues $\gamma = \pm 1$.
Similarly, when $M$ is even, $\psi^{M/2,M}$ corresponds to only $Z_4$ eigenvalues $\gamma = \pm 1$.
The other modes appear in all of eigenstates with eigenvalues $\gamma = \pm 1, \pm i$.
For all the cases with $M=4n, 4n+1, 4n+2, 4n+3$, there are $(n+1)$ independent eigenstates with $\gamma =1$.
It seems convenient to use the  basis such that 
only one of $a_1, a_2, \cdots a_{n+1}$ is non-vanishing in 
 $\sum_{k = 0}^{M-1} a_k \psi^{k,M}$, that is, 
\begin{eqnarray}
& & (a_0,1,0,\cdots, 0,0,a_{n+1},a_{n+2},\cdots, a_{M-1}), \nonumber \\
& & (a_0,0,1,0,\cdots, 0,a_{n+1},a_{n+2},\cdots, a_{M-1}), \nonumber \\
& & ~~~~~~~~~~ \cdots \cdots \cdots   \\
& & (a_0,0,\cdots, 0,1,a_{n+1},a_{n+2},\cdots, a_{M-1}), \nonumber 
\end{eqnarray}
where the other coefficients, $a_0$ and $a_{n+2}, \cdots, a_{M-1}$ are determined by 
eigenvector equations.

Similarly, when there are $m$ independent modes, 
it seems convenient to use the basis such that 
only one of $a_1, a_2, \cdots a_{m}$ is non-vanishing.
The fluxes can be classified as $M=4n, 4n+1, 4n+2, 4n+3$.
For such classes, we show explicitly eigenvectors in the basis $\sum_{k = 0}^{M-1} a_k \psi^{k,M}$ 
in what follows.
As said above, the coefficients $a_k$ other than $a_1, a_2, \cdots a_{m}$ can be written by 
linear combinations of $a_1, a_2, \cdots a_{m}$.
The eigenvectors for $M=$ even are relatively simple, while 
some of eigenvectors for $M=$ odd are written by lengthy linear combinations.
In such cases, we omit write them explicitly and just denote $LP_i(a_1, a_2, \cdots a_{m})$.
At any rate, $LP_i(a_1, a_2, \cdots a_{m})$ can be computed by use of eigenvector equations.

\subsection*{(i)$M=4n+2,\ n \in \mathbb{Z}$}
\noindent
{\bf eigenvalue:$+1$}

$M=2$
\[
\left(
a_0 ,
a_1 ,
\right) \propto \left(
(\sqrt{2}+1) a_1 ,
a_1 
\right) 
\]

$M=6$
\[
\left(
a_0 ,
a_1 ,
a_2 ,
a_3 ,
a_4 ,
a_5
\right) \propto \left(
\frac{1}{2} \left( \sqrt{6} a_1 + (2+\sqrt{6})a_2 \right) ,
a_1 ,
a_2 ,
\frac{1}{2} \left( (2-\sqrt{6})a_1 + \sqrt{6}a_2 \right) ,
a_2 ,
a_1 
\right) 
\]

$M=10$
\[
\left(
a_0 ,\cdots,
a_9
\right) \propto \left(
\frac{1}{2}\left( (1-\sqrt{2}-\sqrt{5}+\sqrt{10})a_1+(2+\sqrt{10})a_2+(-1+\sqrt{2}+\sqrt{5})a_3 \right) ,
\right.
\]
\[
\left.
a_1 ,
a_2 ,
a_3 ,
\frac{1}{2+\sqrt{10}} \left(( -2+2\sqrt{2}+2\sqrt{5}-\sqrt{10})a_1 +(-2-\sqrt{10})a_2 +(2-2\sqrt{2}-2\sqrt{5}+\sqrt{10})a_3 \right) ,
\right.
\]
\[
\left.
\frac{1}{2(2+\sqrt{10})} \left( (-2+5\sqrt{2}+2\sqrt{5}-\sqrt{10})a_1+(-10-2\sqrt{10})a_2+(8-5\sqrt{2}-2\sqrt{5}+\sqrt{10})a_3 \right) ,
\right.
\]
\[
\left.
\frac{1}{2+\sqrt{10}} \left(( -2+2\sqrt{2}+2\sqrt{5}-\sqrt{10})a_1 +(-2-\sqrt{10})a_2 +(2-2\sqrt{2}-2\sqrt{5}+\sqrt{10})a_3 \right)  ,
a_3 ,
a_2 ,
a_1 
\right) 
\]

\ 

\noindent
{\bf eigenvalue:$-1$}

$M=2$
\[
\left(
a_0 ,
a_1 ,
\right) \propto \left(
(-\sqrt{2}+1) a_1 ,
a_1 
\right) 
\]

$M=6$
\[
\left(
a_0 ,
a_1 ,
a_2 ,
a_3 ,
a_4 ,
a_5
\right) \propto \left(
\frac{1}{2} \left( -\sqrt{6} a_1 + (2-\sqrt{6})a_2 \right) ,
a_1 ,
a_2 ,
\frac{1}{2} \left( (2+\sqrt{6})a_1 - \sqrt{6}a_2 \right) ,
a_2 ,
a_1 
\right) 
\]

$M=10$
\[
\left(
a_0 ,\cdots,
a_9
\right) \propto \left(
\frac{1}{2}\left( - (1+\sqrt{2})(-1+\sqrt{5})a_1-(-2+\sqrt{10})a_2+(-1-\sqrt{2}+\sqrt{5})a_3 \right) ,
\right.
\]
\[
\left.
a_1 ,
a_2 ,
a_3 ,
(1+\sqrt{2})a_1+a_2-(1+\sqrt{2})a_3 ,
\frac{1}{2} \left( (1-\sqrt{5})a_1-\sqrt{10}a_2 +(1+\sqrt{5}+\sqrt{10})a_3 \right) ,
\right.
\]
\[
\left.
(1+\sqrt{2})a_1+a_2-(1+\sqrt{2})a_3 ,
a_3 ,
a_2,
a_1 
\right) 
\]

\noindent
{\bf eigenvalue:$+i$}

$M=2$ nothing

$M=6$
\[
\left(
a_0 ,
a_1 ,
a_2 ,
a_3 ,
a_4 ,
a_5
\right) \propto \left(
0 ,
a_1 ,
(-1+\sqrt{2})a_1 ,
0 ,
-(-1+\sqrt{2})a_1 ,
-a_1 
\right) 
\]

$M=10$
\[
\left(
a_0 , \cdots,
a_9
\right) \propto \left(
0 ,
a_1 ,
a_2 ,
\left( -2+\sqrt{5+\sqrt{5}} \right)a_1+ \left( -\sqrt{5}+\sqrt{5-\sqrt{5}} \right)a_2 ,
\right.
\]
\[
\left.
\frac{1}{\sqrt{5-\sqrt{5}}} \left( \left( -5+\sqrt{5}+\sqrt{25-5\sqrt{5}} \right)a_1 +2\left( -\sqrt{5}+\sqrt{5-\sqrt{5}} \right)a_2 \right) ,
0 ,
\right.
\]
\[
\left. 
-\left( \frac{1}{\sqrt{5-\sqrt{5}}} \left( \left( -5+\sqrt{5}+\sqrt{25-5\sqrt{5}} \right)a_1 +2\left( -\sqrt{5}+\sqrt{5-\sqrt{5}} \right)a_2\right) \right) ,
\right.
\]
\[
\left. 
-\left( \left( -2+\sqrt{5+\sqrt{5}} \right)a_1+ \left( -\sqrt{5}+\sqrt{5-\sqrt{5}} \right)a_2 \right) ,
-a_2 ,
-a_1 
\right) 
\]

\noindent
{\bf eigenvalue:$-i$}

$M=2$ nothing

$M=6$
\[
\left(
a_0 ,
a_1 ,
a_2 ,
a_3 ,
a_4 ,
a_5
\right) \propto \left(
0 ,
a_1 ,
(-1-\sqrt{2})a_1 ,
0 ,
-(-1-\sqrt{2})a_1 ,
-a_1 
\right) 
\]

$M=10$
\[
\left(
a_0 ,\cdots,
a_9
\right) \propto \left(
0 ,
a_1 ,
a_2 ,
-\left( 2+\sqrt{5+\sqrt{5}} \right)a_1- \left( \sqrt{5}+\sqrt{5-\sqrt{5}} \right)a_2 ,
\right.
\]
\[
\left. 
\frac{1}{\sqrt{5-\sqrt{5}}} \left( \left( 5-\sqrt{5}+\sqrt{25-5\sqrt{5}} \right)a_1 +2\left( \sqrt{5}+\sqrt{5-\sqrt{5}} \right)a_2 \right) ,
0 ,
\right.
\]
\[
\left.
-\left( \frac{1}{\sqrt{5-\sqrt{5}}} \left( \left( 5-\sqrt{5}+\sqrt{25-5\sqrt{5}} \right)a_1 +2\left( \sqrt{5}+\sqrt{5-\sqrt{5}} \right)a_2 \right) \right) ,
 \right.
\]
\[
\left.
-\left(-\left( 2+\sqrt{5+\sqrt{5}} \right)a_1- \left( \sqrt{5}+\sqrt{5-\sqrt{5}} \right)a_2 \right) ,
-a_2 ,
-a_1 
\right) 
\]

\subsection*{(ii)$M=4n$}
\noindent
{\bf eigenvalue:$+1$}

$M=4$
\[
\left(
a_0 ,
a_1 ,
a_2 ,
a_3 
\right) \propto \left(
2a_1+a_2   ,
a_1 ,
a_2 ,
a_1  
\right) 
\]

$M=8$
\[
\left(
a_0 ,\cdots,
a_7
\right) 
\propto \left(
\frac{1}{\sqrt{2}}a_1+(1+\sqrt{2})a_2+\frac{1}{\sqrt{2}}a_3 ,
a_1 ,
a_2 ,
a_3 ,
\right.
\]
\[
\left.
-\frac{1}{\sqrt{2}}a_1+(1+\sqrt{2})a_2-\frac{1}{\sqrt{2}}a_3 ,
a_3 ,
a_2 ,
a_1 
\right) 
\]

$M=12$
\[
\left(
a_0 ,\cdots,
a_{11}
\right) 
\propto \left(
\frac{1}{2} \left( (3+\sqrt{3})a_2+2\sqrt{3}a_3+(-1+\sqrt{3})a_4 \right) ,
\right.
\]
\[
\left.
a_1 ,
a_2 ,
a_3 ,
a_4 ,
-a_1+\sqrt{3}a_2+2 a_3-\sqrt{3}a_4 ,
\frac{1}{2} \left( (-1+\sqrt{3})a_2-2\sqrt{3}a_3+(3+\sqrt{3})a_4 \right)  ,
\right.
\]
\[
\left.
-a_1+\sqrt{3}a_2+2 a_3-\sqrt{3}a_4   ,
a_4 ,
a_3 ,
a_2 ,
a_1 
\right)
\]

\noindent
{\bf eigenvalue:$-1$}

$M=4$
\[
\left(
a_0 ,
a_1 ,
a_2 ,
a_3 
\right) \propto \left(
-a_1 ,
a_1 ,
a_1 ,
a_1
\right) 
\]

$M=8$
\[
\left(
a_0 ,\cdots,
a_7
\right) 
\propto \left(
-\sqrt{2}a_1+(1-\sqrt{2})a_2 ,
a_1 ,
a_2 ,
a_1 ,
\right.
\]
\[
\left.
\sqrt{2}a_1+(1-\sqrt{2})a_2 ,
a_1 ,
a_2 ,
a_1 
\right) 
\]

$M=12$
\[
\left(
a_0 , \cdots,
a_{11}
\right) 
\propto \left(
(-1-\frac{1}{\sqrt{3}})a_1+(1-\sqrt{3})a_2+(1-\frac{2}{\sqrt{3}})a_3 ,
\right.
\]
\[
\left.
a_1 ,
a_2 ,
a_3 ,
\frac{1}{3} \left( 2\sqrt{3}a_1+3a_2-2\sqrt{3}a_3 \right)  ,
a_1 ,
(-1+\sqrt{3})a_1+(1-\sqrt{3})a_2+a_3  ,
a_1  ,
\right.
\]
\[
\left.
\frac{1}{3} \left( 2\sqrt{3}a_1+3a_2-2\sqrt{3}a_3 \right) ,
a_3 ,
a_2 ,
a_1 
\right)
\]

\noindent
{\bf eigenvalue:$+i$}

$M=4$
\[
\left(
a_0 ,
a_1 ,
a_2 ,
a_3 
\right) \propto \left(
0 ,
a_1 ,
0 ,
-a_1 
\right) 
\]

$M=8$
\[
\left(
a_0 ,\cdots,
a_7
\right) 
\propto \left(
0 ,
a_1 ,
a_2 ,
a_1-\sqrt{2}a_2 ,
0 ,
-(a_1-\sqrt{2}a_2) ,
-a_2 ,
-a_1 
\right) 
\]

$M=12$
\[
\left(
a_0 ,\cdots,
a_{11}
\right) 
\propto \left(
0 ,
a_1 ,
a_2 ,
a_3 ,
2a_1-a_2-(1+\sqrt{3})a_3  ,
\right.
\]
\[
\left.
-a_1-(-1-\sqrt{3})a_3 ,
0  ,
a_1+(-1-\sqrt{3})a_3  ,
-2a_1+a_2+(1+\sqrt{3})a_3  ,
-a_3 ,
-a_2 ,
-a_1 
\right)
\]

\noindent
{\bf eigenvalue:$-i$}

$M=4$ nothing

$M=8$
\[
\left(
a_0 , \cdots,
a_7
\right) \propto \left(
0 ,
a_1 ,
-\sqrt{2}a_1 ,
-a_1 ,
0 ,
a_1 ,
\sqrt{2}a_1 ,
-a_1 
\right) 
\]

$M=12$
\[
\left(
a_0 , \cdots,
a_{11}
\right) \propto \left(
0 ,
a_1 ,
a_2 ,
-(1+\sqrt{3})a_1-(1+\sqrt{3})a_2 ,
\right.
\]
\[
\left.
a_2 ,
a_1+2a_2 ,
0  ,
-a_1-2a_2 ,
-a_2  ,
(1+\sqrt{3})a_1 +(1+\sqrt{3})a_2  ,
-a_2 ,
-a_1 
\right)
\]

\subsection*{(iii)$M=4n+3$}
\noindent
{\bf eigenvalue:$+1$}

$M=3$
\[
\left( a_0 ,a_1, 
a_2 
\right) \propto \left(
(\sqrt{3}+1)a_1 ,
a_1 ,
a_1 ,
\right) 
\]

$M=7$
\[
\left(
a_0 , \cdots ,
a_6 
\right) \propto \left(
LP_0(a_1,a_2) ,
a_1 ,
a_2 ,
LP_3(a_1,a_2) ,
LP_3(a_1,a_2) ,
a_2 ,
a_1 
 \right) 
\]

$M=11$
\[
\left(
a_0 ,\cdots,
a_{10}
\right) 
\propto \left(
LP_0(a_1,a_2,a_3) ,
a_1 ,
a_2 ,
a_3 ,
LP_4(a_1,a_2,a_3) ,
\right.
\]
\[
\left. 
LP_5(a_1,a_2,a_3) , 
LP_5(a_1,a_2,a_3) ,
LP_4(a_1,a_2,a_3) ,
a_3 ,
a_2 ,
a_1 
 \right)
\]

\noindent
{\bf eigenvalue:$-1$}

$M=3$
\[
\left(
a_0 ,
a_1 ,
a_2 
\right) \propto \left(
(-\sqrt{3}+1)a_1 ,
a_1 ,
a_1
\right) 
\]

$M=7$
\[
\left(
a_0 ,\cdots,
a_6 
\right) \propto \left(
LP_0(a_1,a_2) ,
a_1 ,
a_2 ,
LP_3(a_1,a_2) ,
LP_3(a_1,a_2) ,
a_2 ,
a_1 
\right) 
\]

$M=11$
\[
\left(
a_0 , \cdots,
a_{10}
\right)
\propto \left(
LP_0(a_1,a_2,a_3) ,
a_1 ,
a_2 ,
a_3 ,
LP_4(a_1,a_2,a_3) ,
\right.
\]
\[
\left.
LP_5(a_1,a_2,a_3) , 
LP_5(a_1,a_2,a_3) ,
LP_4(a_1,a_2,a_3) ,
a_3 ,
a_2 ,
a_1 
\right)
\]

\noindent
{\bf eigenvalue:$+i$}

$M=3$
\[
\left(
a_0 ,
a_1 ,
a_2 
\right) \propto \left(
0 ,
a_1 ,
-a_1 
\right) 
\]

$M=7$
\[
\left(
a_0 ,\cdots,
a_6 
\right) \propto \left(
0 ,
a_1 ,
a_2 ,
LP_3(a_1,a_2) ,
-LP_3(a_1,a_2) ,
-a_2 ,
-a_1 
\right) 
\]

$M=11$
\[
\left(
a_0 , \cdots,
a_{10}
\right) 
\propto \left(
0 ,
a_1 ,
a_2 ,
a_3 ,
LP_4(a_1,a_2,a_3) ,
LP_5(a_1,a_2,a_3) ,
\right.
\]
\[
\left.
-LP_5(a_1,a_2,a_3) ,
-LP_4(a_1,a_2,a_3) ,
-a_3 ,
-a_2 ,
-a_1 
\right)
\]

\noindent
{\bf eigenvalue:$-i$}

$M=3$ nothing

$M=7$
\[
\left(
a_0 ,\cdots,
a_6 
\right) \propto \left(
0 ,
a_1 ,
LP_2(a_1) ,
LP_3(a_1) ,
-LP_3(a_1) ,
-LP_2(a_1) ,
-a_1 
\right) 
\]

$M=11$
\[
\left(
a_0 ,\cdots,
a_{10}
\right) 
\propto \left(
0 ,
a_1 ,
a_2 ,
LP_3(a_1,a_2) ,
LP_4(a_1,a_2) ,
LP_5(a_1,a_2) ,
\right.
\]
\[
\left.
-LP_5(a_1,a_2) ,
-LP_4(a_1,a_2) ,
-LP_3(a_1,a_2) ,
-a_2 ,
-a_1 
\right)
\]

\subsection*{(iv)$M=4n+1$}
\noindent
{\bf eigenvalue:$+1$}

$M=5$
\[
\left(
a_0 ,
a_1 ,
a_2 ,
a_3 ,
a_4 
\right) \propto \left(
\frac{\sqrt{5}+1}{2} (a_1+a_2) ,
a_1 ,
a_2 ,
a_2 ,
a_1
\right) 
\]

$M=9$
\[
\left(
a_0 , \cdots,
a_8
\right) \propto \left(
LP_0(a_1,a_2,a_3) ,
a_1 ,
a_2 ,
a_3 ,
\right.
\]
\[
\left.
LP_4(a_1,a_2) ,
LP_4(a_1,a_2) ,
a_3 ,
a_2 ,
a_1 
\right) 
\]

\noindent
{\bf eigenvalue:$-1$}

$M=5$
\[
\left(
a_0 ,
a_1 ,
a_2 ,
a_3 ,
a_4 
\right) \propto \left(
(-\sqrt{5}+1)a_1 ,
a_1 ,
a_1 ,
a_1 ,
a_1  
\right) 
\]

$M=9$
\[
\left(
a_0 ,\cdots,
a_8
\right) \propto \left(
LP_0(a_1,a_2) ,
a_1 ,
a_2 ,
LP_3(a_1,a_2) ,
\right.
\]
\[
\left.
LP_4(a_1,a_2) ,
LP_4(a_1,a_2) ,
a_3 ,
a_2 ,
a_1 
\right) 
\]

\noindent
{\bf eigenvalue:$+i$}

$M=5$
\[
\left(
a_0 ,
a_1 ,
a_2 ,
a_3 ,
a_4 
\right) \propto \left(
0 ,
a_1 ,
LP_2(a_1) ,
-LP_2(a_1) ,
-a_1 
\right)  ,
\]
where 
\[
LP_2(a_1) = -\frac{2}{1+\sqrt{5}-\sqrt{2(5+\sqrt{5})}}a_1
\]

$M=9$
\[
\left(
a_0 , \cdots,
a_8
\right) \propto \left(
0 ,
a_1 ,
a_2 ,
LP_3(a_1,a_2) ,
LP_4(a_1,a_2) ,
\right.
\]
\[
\left.
-LP_4(a_1,a_2) ,
-LP_3(a_1,a_2) ,
-a_2 ,
-a_1 
\right) 
\]

\noindent
{\bf eigenvalue:$-i$}

$M=5$
\[
\left(
a_0 ,
a_1 ,
a_2 ,
a_3 ,
a_4 
\right) \propto \left(
0 ,
a_1 ,
LP_2(a_1) ,
-LP_2(a_1) ,
-a_1 
\right)    ,
\]
where 
\[
LP_2(a_1) = -\frac{2}{1+\sqrt{5}+\sqrt{2(5+\sqrt{5})}}a_1
\]

$M=9$
\[
\left(
a_0 ,\cdots,
a_8
\right) \propto \left(
0 ,
a_1 ,
a_2 ,
LP_3(a_1,a_2) ,
LP_4(a_1,a_2) ,
\right.
\]
\[
\left.
-LP_4(a_1,a_2) ,
-LP_3(a_1,a_2) ,
-a_2 ,
-a_1 
\right) 
\]



\end{document}